\documentclass[a4paper, 12 pt] {article}
\usepackage{amsmath}
\usepackage{amsfonts}
\usepackage{amsthm}
\usepackage{amscd} 

\usepackage{graphicx}
\usepackage{epsfig}
\usepackage{color}
\usepackage{graphics}

\newcommand{\half}{{\textstyle \frac{1}{2}}}

\newcommand{\thfourth}{{\textstyle \frac{3}{4}}}

\newcommand{\cvec}{{\bf c}}
\newcommand{\yvec}{{\bf y}}

\newcommand{\evec}{{\bf e}}
\newcommand{\nvec}{{\bf n}}
\newcommand{\pvec}{{\bf p}}
\newcommand{\uvec}{{\bf u}}
\newcommand{\npvec}{{\bf n'}}
\newcommand{\Nvec}{{\bf N}}
\newcommand{\Npvec}{{\bf N'}}
\newcommand{\nuvec}{{\boldsymbol \nu}}

\newcommand{\nuphat}{{\bf \hat \nuvec'}}
\newcommand{\nupphat}{{\bf \hat \nuvec''}}
\newcommand{\svec}{{\bf s}}

\newcommand\sgn{\,{\hbox{\rm sgn}}}

\newcommand{\avec}{{\bf a}}
\newcommand{\bvec}{{\bf b}}
\newcommand{\Cvec}{{\bf C}}
\newcommand{\qvec}{{\bf q}}

\newcommand{\vvec}{{\bf v}}

\newcommand{\xvec}{{\bf x}}
\newcommand{\xivec}{{\boldsymbol \xi}}
\newcommand{\etavec}{{\boldsymbol \eta}}

\newcommand{\fvec}{{\bf f}}
\newcommand{\Hvec}{{\bf H}}
\newcommand{\Fvec}{{\bf  F}}
\newcommand{\zvec}{{\bf z}}
\newcommand{\gvec}{{\bf g}}

\newcommand{\Pbar}{{\bar P}}
\newcommand{\Phat}{{\hat P}}

\newcommand{\Fhat}{{\hat F}}
\newcommand{\Chat}{{\hat C}}
\newcommand{\Bhat}{{\hat B}}
\newcommand{\nhat}{{\bf \hat n}}
\newcommand{\nphat}{{\bf \hat n'}}
\newcommand{\Hhat}{{\bf \hat H}}
\newcommand{\Ehat}{{\hat E}}

\newcommand{\Rcal}{{\cal R}}

\newcommand{\Rr}{{\mathbb R}}

\newcommand{\czero}{{ C^0}}

\newcommand{\inv}{\,\text{\rm Inv}}
\newcommand{\Ical}{{\cal I}}
\newcommand{\epsilonvec}{{\boldsymbol \epsilon}}
\newcommand{\nihat}{{\nhat_\Ical}}
\newcommand{\hhat}{{\bf \hat h}}

\newcommand{\circleddash}{\circ}

\newcommand{\etal}{{\it et al\ }}

\newtheorem{thm}{Theorem}
\newtheorem{lem}{Lemma}[section]
\newtheorem{prop}{Proposition}[section]

\newtheorem{defn}{Definition}[section]

\begin{document}

\title{Classification of unit-vector fields in convex polyhedra with
  tangent boundary conditions}

\author{JM Robbins\thanks{e-mail address: {\tt
  j.robbins@bristol.ac.uk}} \
\& M Zyskin\thanks{e-mail address: {\tt
  m.zyskin@bristol.ac.uk}}\\
School of Mathematics\\ University of Bristol, University Walk, 
Bristol BS8 1TW, UK}

\thispagestyle{empty}

\maketitle






\begin{abstract}
  A unit-vector field $\nvec$ on a convex three-dimensional polyhedron
  $\Pbar$ is tangent if, on the faces of $\Pbar$, $\nvec$ is tangent
  to the faces. A homotopy classification of tangent unit-vector
  fields continuous away from the vertices of $\Pbar$ is given.  The
  classification is determined by certain invariants, namely edge
  orientations (values of $\nvec$ on the edges of $\Pbar$), kink
  numbers (relative winding numbers of $\nvec$ between edges on the
  faces of $\Pbar$), and wrapping numbers (relative degrees of $\nvec$
  on surfaces separating the vertices of $\Pbar$), which are subject
  to certain sum rules.  Another invariant, the trapped area, is
  expressed in terms of these.  One motivation for this study comes
  from liquid crystal physics; tangent unit-vector fields describe the
  orientation of liquid crystals in certain polyhedral cells.
\end{abstract}

\newpage

\section{Introduction}

A unit-vector field $\nvec$ on a convex polyhedron $\Pbar\subset
\Rr^3$ is a map from $\Pbar$ to the unit sphere $S^2\subset \Rr^3$.
$\nvec$ is said to satisfy {\it tangent boundary conditions}, or, more
simply, to be tangent, if, on the faces of $\Pbar$, $\nvec$ is
tangent to the faces.  Tangent boundary conditions imply that, on the
edges of $\Pbar$, $\nvec$ is parallel to the edges, and therefore that
$\nvec$ is necessarily discontinuous at the vertices.  Let
$P \subset \Rr^3$ denote $\Pbar$ without its vertices (thus $\Pbar$ is the
closure of $P$).  Let $\czero(P)$ denote the space of continuous
tangent unit-vector fields on $P$.  We have the usual notion of
homotopic equivalence in $\czero(P)$; two maps $\nvec$, $\npvec \in
\czero(P)$ are homotopic, denoted $\nvec \sim \npvec$, if there exists
a continuous map $\Hvec: P\times [0,1]\rightarrow S^2;
(\xvec,t)\mapsto \Hvec_t(\xvec)$, such that $\Hvec_t$ is tangent and
$\Hvec_0 = \nvec$, $\Hvec_1 = \npvec$.

Here we classify unit-vector fields in $\czero(P)$ up to
homotopy.  The paper is organised as follows.  To a unit-vector field
$\nvec \in \czero(P)$ we associate certain homotopy invariants, which
we call {\it edge orientations}, {\it kink numbers}, and {\it wrapping
  numbers} (Section~\ref{sec: invariants}).  Edge orientations are
just the values of $\nvec$ on the edges of  $P$ (as
noted above, there are two possible values, differing by a sign).
Kink numbers are the integer-valued relative winding numbers of
$\nvec$ between adjacent edges along a face of $P$.
Wrapping numbers are the integer-valued relative degrees of $\nvec$
on planar surfaces which separate one vertex of $P$ 
from the others.  The continuity of $\nvec$ imposes sum rules
on the kink numbers and  wrapping numbers.  In
Section~\ref{sec: representatives} we construct representative maps
for each of the allowed sets of values of the invariants.  In
Section~\ref{sec: classification} we show that an arbitrary map
$\nvec\in\czero(P)$ is homotopic to the reference map with the same
values of the invariants.  One part of the proof, concerning
homotopies on the boundary of $P$, is deferred to Section~\ref{sec: surface}.

We remark that it is the tangent boundary conditions which substantially
determine the classification.  In contrast, continuous unit-vector
fields satisfying {\it fixed} boundary conditions -- for simplicity,
imagine $\nvec$ to be constant on the boundary of $P$ -- are
equivalent to continuous maps of $S^3$ (the unit ball in $\Rr^3$ with
boundary points identified) to $S^2$. As is well known, such maps are
classified by the Hopf invariant.  The absence of a Hopf invariant for
tangent unit-vector fields is due to the vertex discontinuities.

The problem considered here is part of a study of extremals of the
energy functional
\begin{equation}
  \label{eq:energy}
  E = \int_P \sum_{j,k = 1}^3 
  \partial_j n_k \partial_j n_k \,d^3 r
\end{equation}
defined on tangent unit-vector fields in $C^0(P)$ with square-integrable
derivative.  Lower bounds for the energy in terms of the invariants,
along with upper bounds for the case where $P$ is a cube, will be
reported elsewhere~\cite{mrz2003}.

The study of these extremal maps is motivated in part by the study of
liquid crystals in polyhedral cells.  In the continuum limit, the
average local molecular orientation of a uniaxial nematic liquid
crystal may be described by a unit-vector field $\nvec$ (but see
below).  The energy of a configuration $\nvec$ -- the so-called Frank
energy -- reduces, in a certain approximation (the so-called
one-constant approximation), to the expression (\ref{eq:energy}) \cite{degennes}.
Polyhedral liquid crystal cells can be manufactured so that
$\nvec$ is approximately tangent to the cell surfaces.  The homotopy
type of $\nvec$ determines, at least in part, the optical properties
of the liquid crystal, and is relevant to the design of liquid crystal
displays \cite{newtonspiller}.  

In fact, the local orientation of a liquid crystal is only determined
up to a sign, as antipodal orientations are physically equivalent.
Therefore, it is properly described by a director field, a map from
$P$ to the real-projective plane $RP^2$, rather than a unit-vector
field.  However, because $P$ is simply connected, a continuous
director field on $P$ can be lifted to a continuous unit-vector field.
The lifted unit-vector field is determined up to an overall sign.  As
is shown in Section~\ref{sec: invariants}, $+\nvec$ and $-\nvec$
belong to distinct homotopy classes; their kink numbers are the same,
but their edge orientations and wrapping numbers differ by a sign.  By
identifying these pairs of homotopy classes, we obtain a
classification of continuous tangent director fields on $P$.

Twice-differentiable extremals of (\ref{eq:energy}) are examples of
harmonic maps.  Harmonic maps between Riemannian polyhedra have been
studied by Gromov \& Schoen~\cite{gs} and Eells \&
Fuglede~\cite{eellsfug}.  In the case where the target manifold has
nonpositive Riemannian curvature, results concerning the existence,
uniqueness and regularity of solutions of the Euler-Lagrange equations
have been established.  Harmonic unit-vector fields in $\Rr^3$ have
been studied by Brezis \etal~\cite{brezis}, also in connection with
liquid crystals.  The topological classification of liquid crystal
configurations in $\Rr^3$ as well as in domains with smooth boundary
has been extensively discussed -- see, eg, Mermin~\cite{mermin}, de
Gennes and Prost~\cite{degennes}, and Kl\' eman~\cite{kleman}.

We remark that the homotopy classification of tangent unit-vector
fields on $P$ may be regarded as the decomposition of
$\czero(P)$ into its path-connected components with respect to the
{\it compact-open topology}.  The compact-open topology on $\czero(P)$
is generated by sets $[K,U]$, defined for compact $K \subset P$ and
open $U \subset S^2$ by
\begin{equation}
  \label{eq:KU}
  [K,U] = \{\nvec \in \czero(P) |\, \nvec(K) \subset U\}.
\end{equation}
We note that because $P$ is not compact, the compact-open topology on
$\czero(P)$ is distinct from the metric topology on $C^0(P)$,
which is induced by
the metric
\begin{equation}
  \label{eq:eps nbhd}
  d(\nvec,\npvec) = \sup_{\xvec\in P}
    |\,\nvec(\xvec) - \npvec(\xvec)|.
\end{equation}
A path $\Hvec_t\in \czero(P)$ is continuous with respect to the
compact-open topology if and only if $\Hvec_t(\xvec)$ is continuous on
$P \times [0,1]$.  The continuity for $\Hvec_t$ with respect to the metric
topology is a stronger condition; in addition to $\Hvec_t(\xvec)$
being continuous on $P\times [0,1]$, 
$\sup_{x\in P} |\Hvec_t(\xvec) -
\Hvec_{t'}(\xvec)|$ must vanish as $t'$ approaches $t$.

\section{The truncated polyhedron}

Let $\vvec^a$, $a = 1, \ldots, v$, denote the vertices of $P$.  Let
$E^b$, $b = 1,\ldots, e$, denote the edges, and let $F^c$, $c =
1,\ldots, f$, denote the faces.  We regard $E^b$ and $F^c$ as subsets
of $P$.

The truncated polyhedron, denoted $\Phat$, is obtained by cleaving $P$
along planes which separate the vertices from each other.  Explicitly,
let $C^a \subset \Rr^3$ be a plane which separates the vertex
$\vvec^a$ from the vertices $\vvec^{b\ne a}$.  That is, if $\Cvec^a$
denotes a unit normal to $C^a$ and $\cvec^a$ is a point in $C^a$, then
$(\vvec^a -\cvec^a)\cdot \Cvec^a$ and $(\vvec^{b\ne a}-\cvec^a)\cdot
\Cvec^a$ have opposite signs.  For definiteness, we take $\Cvec^a$ to
be outwardly oriented, so that $(\vvec^a -\cvec^a)\cdot \Cvec^a > 0$.
Let $R^a$ denote the closed half-space given by
\begin{equation}
  \label{eq:R^a}
  R^a = \{\xvec \in \Rr^3 | (\xvec -\cvec^a)\cdot \Cvec^a \le 0\}.
\end{equation}
Then the truncated polyhedron $\Phat$ is given by
\begin{equation}
  \label{eq:Phat}
  \Phat = P \cap \left(\cap_{a = 1}^v R^a\right). 
\end{equation}
$\Phat$ is closed and convex.

\begin{figure}
\begin{center}
\input{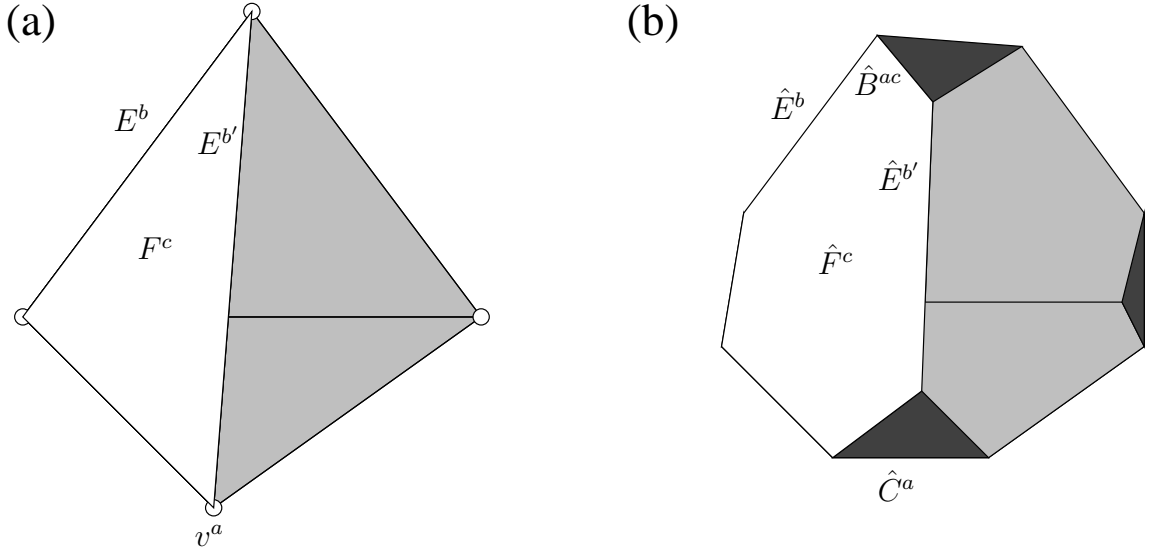}
\caption{(a) The polyhedron $P$ (b) The cleaved polyhedron $\Phat$}
\label{fig: classfig1}
\end{center}
\end{figure}

$\Phat$ has two kinds of faces, which we call {\it cleaved faces} and
{\it truncated faces} (see Fig~\ref{fig: classfig1}).  The cleaved
faces, denoted $\Chat^a$, are given by the intersections of the planes
$C^a$ with $P$.  The truncated faces, denoted $\Fhat^c$, are given by
the intersections of the faces $F^c$ of the original polyhedron 
$P$ with $\cap_{a=1}^v R^a$.

$\Phat$ has two kinds of edges, which we call {\it cleaved edges} and
{\it truncated edges} (see Fig~\ref{fig: classfig1}).  The cleaved
edges, denoted by $\Bhat^{ac}$, are given by the intersections of the
cleaved faces $\Chat^a$ and the truncated faces $\Fhat^c$. The
truncated edges, denoted by $\Ehat^b$, are given by the intersections
of the original edges $E^b$ with $\cap_{a=1}^v R^a$.  The boundaries
of the cleaved faces consist of cleaved edges.  The boundaries of the
truncated faces consist of cleaved edges and truncated edges in
alternation.

We will say that a continuous unit-vector field on $\Phat$ satisfies
tangent boundary conditions if, on the truncated face $\Fhat^c$, the
vector field is tangent to $\Fhat^c$ (note that it need not be tangent
on the cleaved faces).  Let $\czero(\Phat)$ denote the space of
continuous tangent unit-vector fields on $\Phat$.  Given $\nvec \in
\czero(P)$, let $\nhat$ denote its restriction to $\Phat$.  Then
$\nhat \in \czero(\Phat)$.

It turns out that the map $\nvec\mapsto \nhat$ induces a one-to-one
correspondence between homotopy classes of $\czero(P)$ and
$\czero(\Phat)$.
\begin{prop}\label{prop: cp and cphat}
  Given $\nvec$, $\npvec \in \czero(P)$, let $\nhat$, $\nphat \in
  \czero(\Phat)$ denote their restrictions to $\Phat$.  Then $\nvec
  \sim \npvec$ if and only if $\nhat \sim \nphat$.
\end{prop}
\begin{proof}
  Clearly $\nvec \sim \npvec$ implies $\nhat \sim \nphat$.  For the
  converse, we introduce maps $\Nvec$, $\Npvec \in \czero(P)$ which
  coincide with $\nvec$, $\npvec$ on $\Phat$  and are constant
  along rays  in $P-\Phat$ through the vertices.  These rays are of
  the form
\begin{equation}
  \label{eq:rays}
  \xvec^a(r,\yvec^a) = r \yvec^a + (1-r) \vvec^a,
\end{equation}
where $\yvec^a \in \Chat^a$ and $0 < r < 1$.  Every $\xvec \in P -
\Phat$ lies on such a ray and uniquely determines the cleaved face
$\Chat^a$ through which the ray passes as well as  $\yvec^a$ and
$r$.  
Let $\Nvec$ by given by
\begin{equation}
  \label{eq:Nvec}
  \Nvec(\xvec) = 
  \begin{cases}
    \nvec(\xvec),& \xvec  \in \Phat,\\
    \nvec(\yvec^a),& \xvec = \xvec^a(r,\yvec^a).
  \end{cases}
\end{equation}
$\Npvec$ is similarly defined, with $\nvec$ replaced by $\npvec$.  

Assuming that
$\nhat$ and $\nphat$ are homotopic, it follows that $\Nvec$ and
$\Npvec$ homotopic.  Indeed, a homotopy is given by
\begin{equation}
  \label{eq:Hvec}
  \Hvec_t(\xvec) = 
  \begin{cases}
    \Hhat_t(\xvec),& \xvec \in \Phat,\\
    \Hhat_t(\yvec^a),& \xvec = \xvec^a(r,\yvec^a),
  \end{cases}
\end{equation}
where $\Hhat_t$ is a homotopy between $\nhat$ and $\nphat$.

Next we show that $\nvec$ is homotopic to $\Nvec$.  A homotopy
$\Hvec_t$ is given by
\begin{equation}
  \label{eq:homotopy n N}
  \Hvec_t(\xvec) = 
  \begin{cases}
    \nvec(\xvec),& \xvec\in \Phat,\\
    \nvec(\yvec^a),& \xvec =
    \xvec^a(r,\yvec^a) , \ 0 < r < t,\\ 
    \nvec(\xvec^a((r-t)/(1-t),\yvec^a)),& \xvec =
    \xvec^a(r,\yvec^a), \ t \le r < 1.
  \end{cases}
\end{equation}
where $0 \le t \le 1$.
Clearly $\Hvec_0 = \nvec$ and $\Hvec_1 = \Nvec$.  It is
straightforward to verify that $\Hvec_t(\xvec)$ is continuous for
$(\xvec,t) \in P\times [0,1]$ and that it satisfies tangent boundary
conditions.  A similar argument shows that $\npvec$ is homotopic to
$\Npvec$.   Thus we have a chain of equivalences, $\nvec \sim
\Nvec \sim \Npvec \sim \npvec$, which establishes the required result.
\end{proof}

Thus, the homotopy type of tangent unit-vector fields on $P$ is
determined by the homotopy types of their restrictions to the
truncated polyhedron $\Phat$.  Because $\Phat$ is closed, the
classification of the restricted maps is easier to carry out.  For
this reason, we determine homotopy classes of $\czero(\Phat)$ in what
follows.

\section{Invariants}\label{sec: invariants}

Given $\nhat \in \czero(\Phat)$, tangent boundary conditions imply that
its values on the truncated edges $\Ehat^b$ are constant, are tangent to the
edges, and therefore are determined up to a sign.  
\begin{defn}
  \label{def: edge}
The edge orientation $\evec^b(\nhat)$ is the value of $\nhat$ on $\Ehat^b$.
\end{defn}
\noindent The edge orientations are obviously homotopy invariants.  
Under the antipodal map $\nhat \mapsto -\nhat$, the edge orientations
obviously change sign.

Kink numbers are relative winding numbers along cleaved edges.
Let $\zvec^{ac}(t)$, $0\le t \le 1$, denote a
continuous parameterisation of the cleaved edge $\Bhat^{ac}$,
positively oriented with respect to the outward normal, denoted $\Fvec^c$, on
$\Fhat^c$.  Let $  \nhat^{ac}(t) = \nhat(\zvec^{ac}(t))$.
As $\nhat^{ac}(t)$ is tangent to $\Fhat^c$,
its values are
related to $\nhat^{ac}(0)$ by a rotation about $\Fvec^c$, which we
write as
\begin{equation}
  \label{eq:xi^ac}
  \nhat^{ac}(t) = \Rcal(\Fvec^c,\xi^{ac}(t))\cdot \nhat^{ac}(0),
\end{equation}
where $\xi^{ac}(t)$ is the angle of rotation.
We take $\xi^{ac}(t)$ to be
continuous, and fix it uniquely
by taking $\xi^{ac}(0) = 0$.

Let $\eta^{ac}$, where $-\pi < \eta^{ac} <
\pi$, denote the angle (of smallest magnitude) between $\nvec^{ac}(0)$
and $\nvec^{ac}(1)$, so that
\begin{equation}
  \label{eq:eta^ac}
  \nhat^{ac}(1) = \Rcal(\Fvec^c,\eta^{ac})\cdot  \nhat^{ac}(0).
\end{equation}
(Note that since $\nvec^{ac}(0)$ and $\nvec^{ac}(1)$ are parallel to
consecutive truncated edges $\Ehat^b$ and $\Ehat^{b'}$, they cannot be
parallel to each other, so that $\eta^{ac}$ cannot be a multiple of
$\pi$).  From (\ref{eq:xi^ac}) and (\ref{eq:eta^ac}), $\xi^{ac}(1)$
and $\eta^{ac}$ differ by an integer multiple of $2\pi$.  This integer
is the kink number.
\begin{defn}
  \label{def:kink number}
The kink number $k^{ac}(\nhat)$ is given by
\begin{equation}\label{eq: kink def}
    k^{ac}(\nhat) = \frac{1}{2\pi}\left(\xi^{ac}(1) -
  \eta^{ac}\right).
\end{equation}
\end{defn}
The kink number $k^{ac}(\nhat)$ depends continuously on $\nhat$, and
therefore is an integer-valued homotopy invariant on $\czero(\Phat)$.
It may be regarded as the degree (winding number) of the map of
$S^1$ to itself obtained by concatenating $\nhat^{ac}(t)$ 
with a path along which $\nhat^{ac}(1)$ is minimally rotated
back to $\nhat^{ac}(0)$ through their common plane.  

Equations (\ref{eq:xi^ac}) and (\ref{eq:eta^ac}) remain valid if
$\nhat$ is replaced by $-\nhat$.  Therefore,
\begin{equation}
  \label{eq:antipodal kinks}
  k^{ac}(-\nhat) = k^{ac}(\nhat).
\end{equation}

The kink numbers on each truncated face satisfy the following sum rule:
\begin{prop}
  Given $\nhat\in\czero(\Phat)$ and $\Fhat^c$ a truncated face of
  $\Phat$ with outward normal $\Fvec^c$.  Let $q^c(\nhat)$ denote the
  number of pairs of consecutive truncated edges of $\Fhat^c$ on which
  $\nhat$ is oppositely oriented with respect to $\Fvec^c$ (ie,
  $\evec^b(\nhat)\cdot \evec^{b'}(\nhat) < 0$ for consecutive 
  $\Ehat^b$ and $\Ehat^{b'}$).  Then
\begin{equation}
  \label{eq:kink sum rule}
  {\sum_{a}}' k^{ac}(\nhat) = \half q^c(\nhat) - 1,
\end{equation}
where the sum ${\sum_a}'$ is taken over the cleaved edges $\Bhat^{ac}$ of
$\Fhat^c$.
\end{prop}
\begin{proof}
Let $\zvec^c(t)$, $0\le t \le 1$, denote a continuous parameterisation
of 
$\partial \Fhat^c$ (the boundary of $\Fhat^c$), positively oriented with respect to
$\Fvec^c$, with $\zvec^c(1) = \zvec^c(0)$.  Let $\nhat^c(t) =
\nhat(\zvec^c(t))$, and let
\begin{equation}
  \label{eq:Xi}
    \nhat^c(t) = \Rcal(\Fvec^c,\xi^{c}(t))\cdot \nhat^{c}(0),
\end{equation}
where $\xi^c(t)$ is continuous with $\xi^c(0) = 0$. Along the
truncated edges of $\Fhat^c$,
$\xi^c(t)$ is constant.
It follows that $\xi^c(1) = {\sum_{a}}'\xi^{ac}(1)$.
But $\xi^c(1)$ is just $2\pi$ times the winding number of $\nhat$ around
$\partial \Fhat^c$.  Since $\nhat$ is continuous inside $\Fhat^c$, this
winding number vanishes. Therefore
\begin{equation}
  \label{eq:Xi xi}
  {\sum_{a}}'\xi^{ac}(1) = 0.
\end{equation}
Taking the sum ${\sum_{a}}'$ in (\ref{eq: kink def}), 
we conclude that
\begin{equation}
  \label{eq:kink sum rule prelim}
   {\sum_{a}}' k^{ac}(\nhat) = 
-{\sum_{a}}' \frac{1}{2\pi} \eta^{ac}.
\end{equation}

Without loss of generality, we may assume that $F^c$, the face of the
original polyhedron $P$, is a regular polygon ($P$ can be continuously
deformed while remaining convex to make $F^c$ regular).  In this case,
$\nhat^{ac}(0)$ and $\nhat^{ac}(1)$ are parallel to consecutive edges
of a regular polygon. If $\nhat^{ac}(0)$ and $\nhat^{ac}(1)$ are
similarly oriented with respect to $\Fvec^c$, then $\eta^{ac} =
2\pi/m$, where $m$ is the number of sides of $F^c$.  If
they are oppositely oriented, then
$\eta^{ac} = 2\pi/m - \pi$.  Substituting into (\ref{eq:kink sum rule
  prelim}), and noting that there are $m$ terms in the sum, we obtain
the required result (\ref{eq:kink sum rule}).
\end{proof}


Wrapping numbers classify the homotopy type of $\nhat$ on the cleaved
faces $\Chat^a$.  For the explicit definition it will be useful to
introduce  coordinates on $\Chat^a$.  Let $\zvec^a(\phi)$ denote a
piecewise-differentiable, $2\pi$-periodic parameterisation of
$\partial \Chat^a$, positively oriented with respect to the outward
normal, denoted $\Cvec^a$, on $\Chat^a$.  Let $\cvec^a$ be a point in the
interior of $\Chat^a$, and let
\begin{equation}
  \label{eq:x(rho, yvec)}
  \yvec^a(\rho,\phi) = \rho\zvec^a(\phi) + (1-\rho)\cvec^a,
\end{equation}
where $0\le\rho\le1$.  

To a map $\nhat\in\czero(\Phat)$, we associate a continuous map $\nuvec^a$ from
$D^2$, the unit two-disk, to $S^2$, given by   
\begin{equation}
  \label{eq:nhat^a}
  \nuvec^a(\rho,\phi) = \nhat(\yvec^a(\rho,\phi)).
\end{equation}
We construct another continuous map $\nuvec_0^a:D^2\rightarrow S^2$ as
follows.  On the boundary of the disk, $\nuvec_0^a$ is taken to
coincide with $\nuvec^a$.  Along radial lines from the boundary to the
centre of the disk, $\nuvec_0^a$ is taken to trace out the shortest
geodesic from its value on the boundary to a fixed value, which we
denote by $-\svec$. (In what follows, we sometimes regard $\svec$ as the
south pole of $S^2$, and $-\svec$ as the north pole.)  Explicitly, let
$\gvec_\rho(-\svec,\avec)$, where $0 \le \rho \le 1$, denote the
shortest geodesic arc from $-\svec$ to $\avec$, where the parameter
$\rho$ is proportional to arclength.  Then $\nuvec_0^a: D^2\rightarrow
S^2$ is given by
\begin{equation}
  \label{eq:nhat^a_0}
  \nuvec_0^a(\rho,\phi) = \gvec_\rho(-\svec, \nhat(\zvec^a(\phi))).
\end{equation}
(\ref{eq:nhat^a_0}) is well defined provided that the boundary values of $\nhat$ are
not antipodal to $-\svec$, ie $\nhat \ne \svec$ on $\partial \Chat^a$.
Since, on $\partial \Chat^a$, $\nhat$ is tangent to a truncated face,
this condition is satisfied provided that 
\begin{equation}
  \label{eq:svec condition}
  \svec\cdot \Fvec^c \ne 0, \quad c = 1,\ldots, f.
\end{equation}
From now on, we assume $\svec$ is chosen to satisfy (\ref{eq:svec condition}).
Note that, by construction, $\svec$ is not in the image of $\nuvec^a_0$.

Given two maps $\nuvec^a, \nuvec^a_0:D^2 \rightarrow S^2$ which
coincide on
$\partial D^2$, we may glue them on the boundary to get a continuous
map on $S^2$, which we denote by $\nuvec^a\circleddash
\nuvec^a_0$.
Explicitly, $\nuvec^a\circleddash \nuvec^a_0: S^2 \rightarrow S^2$ is given by
\begin{equation}
  \label{eq:nuvec^a}
 (\nuvec^a\circleddash \nuvec^a_0)(x,y,z) = 
  \begin{cases}
    \nuvec^a(\rho,\phi),&z\ge 0,\\
    \nuvec^a_0(\rho,\phi),&z < 0,
  \end{cases}
\end{equation}
where $(\rho,\phi)$ are the polar coordinates of $(x,y)$.  The
wrapping number is the degree of this map.
\begin{defn}
  \label{defn: wrapping number}
The wrapping number $w^a(\nhat)$ is given by
\begin{equation}\label{eq: wrapping def}
  w^a(\nhat) = \deg (\nuvec^a\circleddash\nuvec_0^a).
\end{equation}
\end{defn}
\noindent The wrapping number depends continuously on $\nhat$ (since
$\nuvec^a$ and $\nuvec_0^a$ do), and therefore is
a homotopy invariant.

For $\nhat \in C^1(\Phat)$ (ie, $\nhat$ is continuously
differentiable on $\Phat$), we derive an integral formula for the
wrapping number.  We take $\zvec^a(\phi)$ to be
piecewise-$C^1$, so that the derivative of $\nuvec^a\circleddash\nuvec_0^a$ is piecewise
continuous.  Then
\begin{equation}
  \label{eq:degree phivec integral}
  \deg (\nuvec^a\circleddash\nuvec_0^a) = \frac{1}{4\pi}\int_{S^2} 
\left(\nuvec^a\circleddash\nuvec_0^a\right)^*\,\omega,
\end{equation}
where $\omega$ is the rotationally invariant area-form on $S^2$,
normalised so that $\int_{S^2} \omega = 4\pi$, and
$\left(\nuvec^a\circleddash\nuvec_0^a\right)^*$ denotes the pull-back.
From (\ref{eq:nuvec^a}),
\begin{equation}
  \label{eq:wa third prelim}
  w^a(\nhat) = \deg (\nuvec^a\circleddash\nuvec_0^a) =
\frac{1}{4\pi}\int_{D^2} \left({\nuvec^a}^*\,\omega - {\nuvec^a_0}^*\, \omega\right) =
\frac{1}{4\pi}\int_{\Chat^a} \nhat^*\, \omega - \frac{1}{4\pi}\int_{D^2} {\nuvec^a_0}^*\, \omega.
\end{equation}

By construction, $\nuvec^a_0$ takes values in $\{S^2 - \svec\}$ (the
two-sphere with the point $\svec$ removed).  Let $\gamma$ denote a
one-form on $\{S^2 -\svec\}$ for which 
\begin{equation}
  \label{eq:d gamma}
  d\gamma = \omega \text{\ on } \{S^2 - \svec\}.
\end{equation}
For example, we may take $\gamma = (1 - \cos \alpha)d\beta$,
where $(\alpha,\beta)$ are spherical polar coordinates on $S^2$ with
south pole at $\svec$.  Applying Stokes' theorem to the second
integral in (\ref{eq:wa third prelim}), we get
\begin{equation}
  \label{eq:wa third}
  w^a(\nhat) = \frac{1}{4\pi}\left(\int_{\Chat^a} \nhat^*\, \omega - \int_{\partial \Chat^a} \nhat^*\, \gamma\right).
\end{equation} 
From (\ref{eq:wa third}), it is clear that wrapping numbers change sign
under the antipodal map $\nhat \rightarrow -\nhat$,
 \begin{equation}
  \label{eq:winding signs}
   w^a(-\nhat) = -w^a(\nhat).
 \end{equation} 
 In fact, (\ref{eq:winding signs}) holds for all maps in
 $\czero(\Phat)$, since any map in $\czero(\Phat)$ is homotopic to a
 $C^1$-map in $\czero(\Phat)$.

If $\svec$ is a regular value of
$\nhat$ on $\Chat^a$ -- that is, if the derivative $ d\nhat$ 
restricted to $\Chat^a$ has full rank on $\nhat^{-1}(\svec)$ -- then
$\nhat^{-1}(\svec)$ is a finite set, and we
can express the wrapping number as a signed count of preimages
$\yvec^a_*$ of
$\svec$ on $\Chat^a$.   
We have that
\begin{equation}
  \label{eq:small neighbourhood}
  \Chat^a = \left(\Chat^a - 
\sum_{\yvec^a_*}
U_\epsilon (\yvec^a_* )\right) + 
\sum_{\yvec^a_*}
U_\epsilon (\yvec^a_*),
\end{equation}
where $U_\epsilon(\yvec^a_*)$ is an $\epsilon$-neighbourhood of
$\yvec^a_*$. Substituting (\ref{eq:small neighbourhood}) into the integral formula (\ref{eq:wa third}),
the contribution from the first term in (\ref{eq:small neighbourhood})
vanishes due to Stokes' theorem, while each neighbourhood
$U_\epsilon(\yvec^a_*)$ in the second term contributes $\sgn \det
 d\nhat(\yvec^a_*)$, where the determinant is computed with respect to positively
oriented coordinates on $\Chat^a$ and $S^2$.  Then
\begin{equation}
  \label{eq:w^a second}
  w^a(\nhat) = \sum_{\yvec^a_*} 
\sgn \det  d\nhat(\yvec^a_*).
\end{equation}

Next we use (\ref{eq:wa third}) to show that the sum
of the wrapping numbers vanishes.  
\begin{prop}
  \label{prop: wrap sum} 
Given $\nhat\in\czero(\Phat)$, 
\begin{equation}
\label{eq: wrap sum}
\sum_{a=1}^v w^a(\nhat) = 0.
\end{equation}
\end{prop}
\begin{proof}
  We have that
\begin{equation}
  \label{eq:sum of w^a}
  \sum_{a=1}^v w^a(\nhat) = \sum_{a=1}^v \int_{\Chat^a} \nhat^*\,\omega - 
  \sum_{a=1}^v \int_{\partial \Chat^a} \nhat^*\,\gamma.
\end{equation}
The boundary of the truncated polyhedron $\Phat$ is given by 
\begin{equation}
  \label{eq:boundary}
  \partial \Phat = \sum_{a = 1}^v  \Chat^a + \sum_{c=1}^f  \Fhat^c.
\end{equation}
Since $\partial(\partial \Phat) = 0$,
\begin{equation}
  \label{eq:sides cancelling}
 \sum_{a = 1}^v  \partial \Chat^a + \sum_{c=1}^f  \partial \Fhat^c = 0.
\end{equation}
The second integral in (\ref{eq:sum of w^a}) may then be rewritten as
\begin{equation}
  \label{eq:second integral}
   \sum_{a=1}^v \int_{\partial \Chat^a} \nhat^*\,\gamma = 
-\sum_{c=1}^f \int_{\partial
  \Fhat^c} \nhat^*\,\gamma = -\sum_{c=1}^f \int_{\Fhat^c} \nhat^*\,\omega,
\end{equation}
where in the second equality we have used Stokes' theorem and (\ref{eq:d gamma}) (this is
justified since $\nhat \ne \svec$ on $\Fhat^c$).  Substituting
(\ref{eq:second integral}) into (\ref{eq:sum of w^a}), we get
\begin{equation}
  \label{eq:wrap sum 2}
  \sum_{a=1}^v w^a(\nhat) = \left(\sum_{a=1}^v\int_{\Chat^a} +
    \sum_{c=1}^f \int_{\Fhat^c}\right) \nhat^*\, \omega = 
  \int_{\partial \Phat} \nhat^*\, \omega.
\end{equation}
Since $\omega$ is closed, the last expression vanishes.  Therefore
\begin{equation}
  \label{eq:wrap sum rule}
  \sum_{a=1}^v w^a(\nhat) = 0.
\end{equation}
This result applies to all maps in $\czero(\Phat)$, as every map
in $\czero(\Phat)$ is homotopic to a $C^1$-map in $\czero(\Phat)$.
\end{proof}

The wrapping number depends on the
choice of $\svec \in S^2$.  For $\nhat\in C^1(\Phat)$, 
an alternative, convention-independent
invariant is the real-valued quantity
\begin{equation}
  \label{eq:trapped area}
  \Omega^a(\nhat) = \int_{\Chat^a} \nhat^*\, \omega = 
  4\pi w^a(\nhat)  + \int_{\partial\Chat^a} \nhat^*\,\gamma.
\end{equation}
We call $\Omega^a$ the {\it trapped area} at the vertex $a$.  
It plays a central role in estimates of lower bounds for
the energy (\ref{eq:energy}) (\cite{brezis}, \cite{mrz2003}).

The second term in the expression (\ref{eq:trapped area}) for the
trapped area can be expressed 
in terms of the kink numbers and edge orientations, as we now show. We have that
\begin{equation}
  \label{eq:n*(partial Ca)}
  \nhat(\partial \Chat^a) = {\sum_{c}}' k^{ac} S^{1c} + K^a,
\end{equation}
where, in the first term, $S^{1c}$ denotes the unit circle in $S^2$
normal to $\Fvec^c$, positively oriented with respect to $\Fvec^c$,
and the sum ${\sum_c}'$ is taken over the cleaved edges $\Bhat^{ac}$
of $\partial \Chat^a$.  From (\ref{eq:d gamma}), the integral of
$\gamma$ around $S^{1c}$ is given by $-2\pi \sgn (\Fvec^c
\cdot\svec)$.  The second term in (\ref{eq:n*(partial Ca)}), $K^a$, is
the geodesic polygon in $S^2$ with vertices
$\evec^{b_1}(\nhat),\ldots\evec^{b_m}(\nhat)$, where the indices
$b_r$ label the truncated edges $\Ehat^{b_1}, \ldots,
\Ehat^{b_m}$ of $\partial \Chat^a$, consecutively ordered with respect
to the outward normal.  Suppose first that $K^a$ has just three
vertices, which we denote $\avec$, $\bvec$ and $\cvec$ for
convenience.  From (\ref{eq:d gamma}),
\begin{equation}
  \label{eq:gamma triangle}
  \int_{K^a} \gamma =
  A(\avec,\bvec,\cvec) -  4\pi \sigma(\avec,\bvec,\cvec),
\end{equation}
where $A(\avec,\bvec,\cvec)$ is the oriented area of
$K^a$, with the interior of $K^a$ chosen so that $|A(\avec,\bvec,\cvec)| < 2\pi$,
and $\sigma(\avec,\bvec,\cvec) =
\pm 1, 0$ according to whether $\svec$ is outside $K^a$ (in which case $\sigma =
0$) or inside $K^a$ (in which case $\sigma$ is the orientation of $\partial K^a$ about $\svec$).
Explicitly, $A(\avec,\bvec,\cvec)$ is given by 
\begin{equation}
  \label{eq:triangle area}
  A(\avec,\bvec,\cvec) = 2\arg ((1 + \avec\cdot\bvec +
  \bvec\cdot\cvec + \cvec\cdot\avec) + i (\avec\times\bvec)\cdot\cvec),
\end{equation}
where $\arg$ is taken between $-\pi$ and $\pi$ ((\ref{eq:triangle
  area}) is equivalent to the standard expression $\alpha + \beta
  + \gamma - \pi$ for the area of a unit spherical triangle with
  interior angles $\alpha$, $\beta$ and $\gamma$.)
$\sigma(\avec,\bvec,\cvec)$ is given by
\begin{equation}
  \label{eq:sigma}  \sigma(\avec,\bvec,\cvec) = 
  \begin{cases}
    \sgn ((\avec\times\bvec)\cdot \svec),& \svec \in K^a,\\
    0,& \svec \notin K^a.
  \end{cases}
\end{equation}
In fact, $\svec \in K^a$ if and only if $(\avec\times\bvec)\cdot \svec$,
$(\bvec\times\cvec)\cdot \svec$ and $(\cvec\times\avec)\cdot \svec$
all have the same sign.
If $K^a$ has $m > 3$ vertices, we may represent it as a sum of geodesic triangles $K^a_j$
with vertices
$\evec^{b_1}(\nhat)$, $\evec^{b_j}(\nhat)$, $\evec^{b_{j+1}}(\nhat)$,
with $2 \le j \le m -1$.  

These considerations are summarised in the following:
\begin{prop}
  \label{prop:trapped area}
  Given a cleaved face $\Chat^a$ with truncated edges
  $\Ehat^{b_1},\ldots E^{b_m}$ consecutively ordered with respect to
  the outward orientation.  The trapped area (\ref{eq:trapped area})
  is given by
\begin{equation*}
\Omega^a = 
4\pi w^a  -  2\pi{\sum_{c}}' 
\sgn(\Fvec^c\cdot\svec) k^{ac}
+  \sum_{j=2}^{m-1} \left(A(\evec^{b_1},\evec^{b_j},\evec^{b_{j+1}}) - 4\pi 
\sigma(\evec^{b_1},\evec^{b_j},\evec^{b_{j+1}})\right),
\end{equation*}
where the sum $\sum_{c}'$ is taken over the cleaved edges $\Bhat^{ac}$ of
$\Chat^a$, and $A$ and $\sigma$ are given by (\ref{eq:triangle area})
and (\ref{eq:sigma}) respectively.
\end{prop}


\section{Representatives}\label{sec: representatives}

Let 
\begin{equation}
  \label{eq:Inv}
  \inv = \{\evec^b, k^{ac}, w^a\}
\end{equation}
denote the set of homotopy
invariants on $\czero(\Phat)$ defined in Section~\ref{sec:
  invariants}.  Let $\Ical = (\epsilonvec^b, \kappa^{ac}, \omega^a\}$
denote a set of values of $\inv$ which satisfies the 
sum rules (\ref{eq:kink sum rule}) and (\ref{eq:wrap sum rule}).
In what follows, we construct a representative map $\nihat \in
\czero(\Phat)$ for which
\begin{equation}
  \label{eq:nihat}
  \inv(\nihat) = \Ical.
\end{equation}

We first define $\nihat$ on the edges of $\Phat$.  On the truncated
edges, $\nihat$ is determined by the edge orientations, $\epsilonvec^b$.
\begin{equation}
  \label{eq:nihat on truncated edges}
  \nihat(\xvec) = \epsilonvec^b,\ \xvec \in \Ehat^b.
\end{equation}
On the cleaved edges, $\nihat$ is determined up to homotopy by the
edge orientations and the kink numbers, $\kappa^{ac}$.  Let
$\zvec^{ac}(t)$, $0\le t\le 1$, denote a parameterisation of
$\Bhat^{ac}$, positively oriented with respect to $\Fvec^c$.  Let the
endpoints $\zvec^{ac}(0)$ and $\zvec^{ac}(1)$ lie on consecutive
truncated edges $\Ehat^b$ and $\Ehat^{b'}$ respectively.  Let $\eta^{ac}\in
(-\pi,\pi)$ denote the angle from $\epsilonvec^b$ to $\epsilonvec^{b'}$, as in (\ref{eq:eta^ac}).  Then on
$\Bhat^{ac}$, we take $\nihat$ to be given by
\begin{equation}
  \label{eq:nihat on cleaved edges}
  \nihat(\zvec^{ac}(t)) = \Rcal(\Fvec^c,(\eta^{ac} +
  2\pi\kappa^{ac})t)\cdot \epsilonvec^b.
\end{equation}

To extend $\nihat$ to the faces of $\Phat$, it is
convenient to introduce polygonal-polar coordinates.  Let $\fvec^c$ be
a point in the interior of the truncated face $\Fhat^c$. We
parameterise $\Fhat^c$ by
\begin{equation}
  \label{eq:param of Fhatc}
  \yvec^c(\rho,\zvec^c) = \rho \zvec^c + (1-\rho)\fvec^c,
\end{equation}
where $0\le \rho \le 1$ and $\zvec^c \in \partial \Fhat^c$.
By a radial chord, we mean the segment obtained by taking
$\zvec^c$ fixed in (\ref{eq:param of Fhatc}), and letting $\rho$ vary
between $0$ and $1$.
Similarly, let $\cvec^a$ be a point in the interior of the cleaved
face $\Chat^a$. 
We
parameterise $\Chat^a$ by
\begin{equation}
  \label{eq:param of Chata}
  \yvec^a(\rho,\zvec^a) = \rho \zvec^c + (1-\rho)\cvec^a,
\end{equation}
Radial chords on $\Chat^a$ are defined as for $\Fhat^c$.

On $\Fhat^c$, we define $\nihat$ along radial chords by contracting its
boundary values to a constant.  Explicitly, we note that
(\ref{eq:nihat on truncated edges}) and (\ref{eq:nihat on cleaved
  edges}) determine $\nihat$ on $\partial \Fhat^c$.  We regard
$\nihat$ on $\partial \Fhat^c$ as a continuous map of $S^1$ to itself
(the image lies in $S^{1c}$, the great circle orthogonal to $\Fhat^c$).  Since the kink numbers $\kappa^{ac}$
satisfy the sum rule (\ref{eq:kink sum rule}), this map has zero
winding number, and therefore is contractible.  That is, there exists
a continuous unit-vector field $\hhat_t(\zvec^c)$ tangent to $\Fhat^c$
such that ${\hhat}^c_0(\zvec^c) = \nihat(\zvec^c)$ and ${\hhat^c}_1$
is constant. Let
\begin{equation}
  \label{eq:nihat on F^c}
  \nihat(\yvec^c(\rho,\zvec^c)) = {\hhat^c}_\rho(\zvec^c).
\end{equation}

On $\Chat^a$, we note that (\ref{eq:nihat on cleaved edges})
determines the values of $\nihat$ on $\partial \Chat^{a}$, where
$\rho = 1$.
We define $\nihat$ for $\half \le \rho < 1$ by
contracting its boundary values along shortest geodesics
on $S^2$ to $-\svec$.
\begin{equation}
  \label{eq:nihat, rho > half}
  \nihat(\yvec^a(\rho,\zvec^a)) = g_{2\rho-1}(-\svec,
  \nihat(\yvec^a(1,\zvec^a))),\ \ \half \le \rho < 1,
\end{equation}
where $g_\tau(-\svec,\avec)$, $0 \le \tau \le 1$, denotes the shortest
geodesic from $-\svec$ to $\avec$ (as in (\ref{eq:nhat^a_0})).  
For $\rho \le \half$, we insert a covering of $S^2$ with multiplicity
given by the wrapping number $\omega^a$.  Explicitly, let $\zvec^a(\phi)$ be a
$2\pi$-periodic parameterisation of $\partial \Chat^a$, and let 
\begin{equation}
  \label{eq:nihat, rho < half}
  \nihat(\yvec^a(\rho,\zvec^a(\phi))) = \sin 2\pi \rho \cos \omega^a
  \phi\, \xivec + \sin 2\pi \rho \sin \omega^a \phi\, \etavec + \cos 2\pi \rho\,\svec, \  0 \le \rho
  < \half,
\end{equation}
where $\xivec$ and $\etavec$ are orthonormal vectors in the plane
perpendicular to $\svec$ with $\xivec \times \etavec = -\svec$.  Let
$(\alpha,\beta)$ denote polar coordinates on $S^2$ with south pole at
$\svec$.  Identifying $S^2$ with the region $\rho \le \half$ on
$\Chat^{a}$ via
$\rho = (\pi -
\alpha)/2\pi$, $\zvec^a = \zvec^a(\beta)$, then (\ref{eq:nihat, rho <
  half}) corresponds to the $S^2$-map $(\alpha,\beta)\mapsto
(\alpha,\omega^a\beta)$, which has degree $\omega^a$.  It is readily
verified from (\ref{eq: wrapping def}) that $w^a(\nihat) = \omega^a$.

We extend $\nihat$ to the interior of $\Phat$ along radial lines by
contracting its boundary values to a constant.  Explicitly, we note
that (\ref{eq:nihat on F^c}), (\ref{eq:nihat, rho > half}) and
(\ref{eq:nihat, rho < half}) 
determine
$\nihat$ on $\partial \Phat$.  From (\ref{eq:wrap sum 2}), the
integral of $\nihat^*\,\omega$ over $\partial \Phat$ is given by the
sum of the wrapping numbers $\omega^a$.  By assumption, this sum
vanishes, so that
\begin{equation}
  \label{eq:nihat on Phag}
  \int_{\partial \Phat} \nihat^*\, \omega = 0
\end{equation}
(we can take $\nihat$ to be piecewise-differentiable on
$\partial\Phat$, so that 
$\nihat^*\, \omega$ is piecewise-continuous).  Regarding
$\partial \Phat$ as a topological two-sphere, we may regard $\nihat$
on $\partial \Phat$
as a degree-zero map on $S^2$.  There exists a contraction to a
constant map. 
Let $\hhat_t:\partial \Phat\rightarrow S^2$, where $0\le t \le 1$, be such a
contraction, ie $\hhat_t$ is 
continuous,  $\hhat_0 = \nihat$ and $\hhat_1 = \svec$, constant.
Let
$\pvec$ be a point in the interior of $\Phat$, and let
\begin{equation}
  \label{eq:xvec}
  \xvec(r,\yvec) = r\yvec + (1-r)\pvec,
\end{equation}
where $0 \le r \le 1$.  Then we define $\nihat$ in $\Phat$ by
\begin{equation}
  \label{eq:nihat in Phat}
  \nihat(\xvec(r,\yvec))) = \hhat_r(\yvec).
\end{equation}

Let $\cvec^{a_*}$ denote the interior point of the cleaved face
$\Chat^{a_*}$.  Setting $\rho = 0$ in (\ref{eq:nihat, rho < half}), we
see that $\nihat(\cvec^{a_*}) = \svec$.  Without loss of generality,
and for future convenience, we
choose the homotopy $\hhat_t$ so that $\hhat_t(\cvec^{a_*}) =
\svec$ for all $0 \le t \le 1$. 
Therefore, from
(\ref{eq:nihat in Phat}),
\begin{equation}
  \label{eq:marked point}
  \nihat(\xvec(\rho,\cvec^{a_*})) = \svec.
\end{equation}

We note that the construction of $\nihat$ is not completely explicit,
in that we make use of the contractibility of degree-zero maps on
$S^1$ and $S^2$ without specifying these contractions explicitly.  An
explicit prescription for these contractions (which is valid for all
$S^n$) is described by, eg,  Whitehead~\cite{whitehead} (of course, for
$S^1$, the contraction is easily constructed).

\section{Classification}\label{sec: classification}

Our main result is that the invariants, $\inv$, classify maps in
$\czero(\Phat)$ up to homotopy.
\begin{thm}
  \label{thm: classify}
  Let $\nhat, \nphat \in \czero(\Phat)$.  Then $\nhat \sim \nphat$
  if and only if
$\inv(\nhat) = \inv(\nphat)$.
\end{thm}

\begin{proof}
  Since $\inv(\nhat)$ is homotopy invariant, it is clear that $\nhat \sim
  \nphat$ only if $\inv(\nhat) = \inv(\nphat)$.  For the converse, it
  suffices to show that $\nhat$ is homotopic to the representative map
  $\nihat$, where $\Ical = \inv(\nhat)$.  
  
  It will be convenient to use the polyhedral-polar coordinates
  $\xvec(r,\yvec)$ on $\Phat$ given by (\ref{eq:xvec}), where $0\le r
  \le 1$ and $\yvec\in\partial \Phat$.  The sets $r = \text{constant}$
  interpolate between the boundary $\partial \Phat$ ($r = 1$) and the
  interior point $\pvec$ ($r = 0$).  Let $\Phat(a,b)$ denote the
  polyhedral shell $a \le r \le b$.  With an abuse of notation,
  we shall sometimes write, for the sake of brevity, $\nhat(r,\yvec)$,
  rather than $\nhat(\xvec(r,\yvec))$, and similarly for other maps in $\czero(\Phat)$.
  
  To show that $\nhat \sim \nihat$, we argue as follows.  First, we
  deform $\nhat$ to a map $\nhat_1$ which coincides with a radially
  scaled copy of $\nihat$ on the outer shell $\Phat(\half,1)$ and
  which is constant, equal to $\svec$, on the inner shell
  $\Phat(\epsilon,\half)$, where $\epsilon > 0$ is specified below.
  The dependence of $\nhat_1$ on the original map
  $\nhat$ is confined to the polyhedral bubble $\Phat(0,\epsilon)$.
  Then, we create a radial channel through the outer shell, inside of
  which the map is made to be constant, equal to $\svec$.  The
  polyhedral bubble is made to evaporate through this channel.
  The channel is then removed, leaving a map which is a radially scaled
  copy of $\nihat$ on $\Phat(\half,1)$ and which is constant, equal to $\svec$,
  on $\Phat(0,\half)$.  A final rescaling produces $\nihat$.  A
  schematic description of these deformations is shown in Fig~\ref{fig: classfig2}.  Details
  of the argument follow below.
  
  Without loss of generality, we may assume that $\nhat$ coincides
  with $\nihat$ on $\partial \Phat$; this is demonstrated in the
  following section (see Prop~\ref{prop: surface homotopy}).  Then for
  any $0<\epsilon <\half$, $\nhat$ is homotopic to a map $\nhat_1\in
  \czero(\Phat)$ given by
\begin{equation}
  \label{eq:nhat_1}
  \nhat_1(r,\yvec) = 
  \begin{cases}
    \nihat(2r-1,\yvec),&\half\le r\le 1,\\
    \svec,&\epsilon\le r < \half
  \end{cases}
\end{equation}
for $\epsilon \le r \le 1$.
Note that, from (\ref{eq:marked point}), $\nihat(0,\yvec) = \svec$, so
that $\nhat_1$ is continuous at $r = \half$.  For $r < \epsilon$,
$\nhat_1$ is given by
\begin{equation}
  \label{eq:nhat_1 inside}
  \nhat_1(r,\yvec) = 
\begin{cases}
  \nihat(2(\epsilon-r)/\epsilon,\yvec),&\half \epsilon \le r <
    \epsilon,\\
    \nhat(2r/\epsilon, \yvec), & 0 \le r < \half \epsilon.
  \end{cases}
\end{equation}
See Fig~2(b).  In fact, the particular form for $r \le \epsilon$ will not concern us
in what follows.
A homotopy between $\nihat$ and $\nhat_1$ is given by
\begin{equation}
  \label{eq:n->n1 homotopy}
  \Hvec_t(r,\yvec) = 
  \begin{cases}
    \nihat(\sigma(r),\yvec),&1-\half t\le r \le 1,\\
     \nihat(1-t,\yvec),&1-(1-\epsilon)t\le r < 1-\half t,\\
    \nihat(\tau_t(r),\yvec),&1-(1-\half\epsilon)t\le r <
    1-(1-\epsilon)t,\\
    \nhat(\upsilon_t(r),\yvec),&r < 1-(1-\epsilon/2)t,
  \end{cases}
\end{equation}
where
\begin{eqnarray}
  \label{eq:sigma,tau,upsilon}
  \sigma(r) &=& 2r-1,\nonumber\\
  \tau_t(r) &=& 1 +2((1-r) - (1-\half\epsilon)t)/\epsilon,\nonumber\\
  \upsilon_t(r) &=& r/(1 - (1-\half\epsilon)t).
\end{eqnarray}

\begin{figure}
\begin{center}
\input{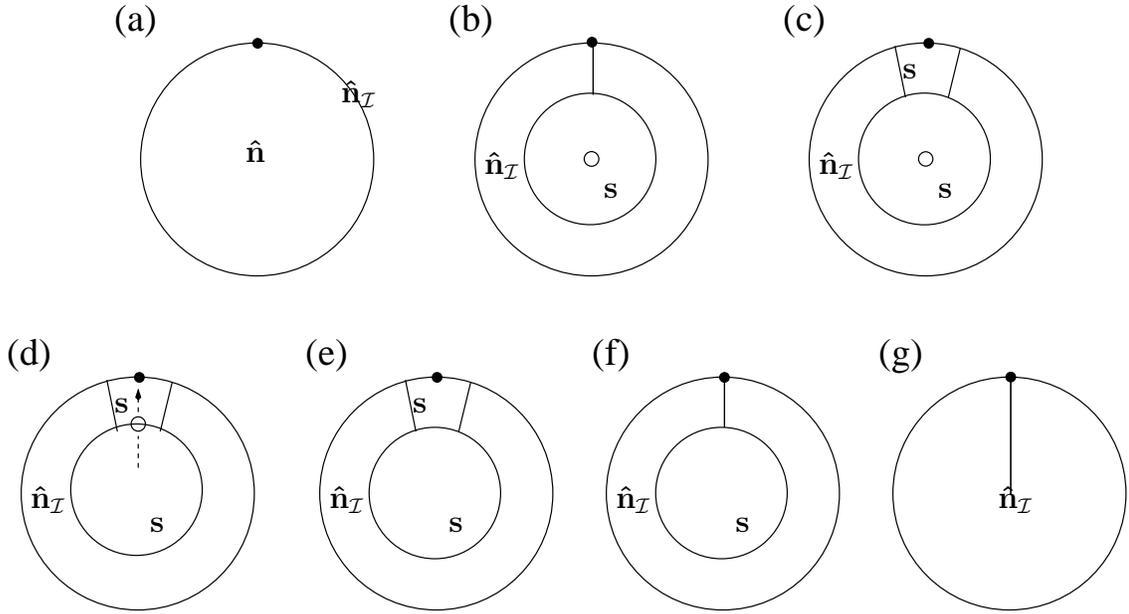}
\caption{Homotopy from $\nhat$ to $\nihat$.  Polyhedral shells
  $\Phat(a,b)$ are represented schematically as spherical shells. (a)
  $\nhat$ coincides with $\nihat$ on $\partial \Phat$. The marked
  point is $\cvec^{a_*}$, where $\nihat = \svec$. (b) $\nhat_1$.  Note
  that $\nhat_1 = \svec$ along the outer half of the ray from
  $\cvec^{a_*}$ to the centre. (c) $\nhat_2$ is equal to $\svec$ in the
  channel. (d) The polyhedral bubble, $P(0,\epsilon)$, is floated
  through the channel. (e) $\nhat_3$ (f) The channel is removed to
  obtain $\nhat_4$. (g) $\nihat$}
\label{fig: classfig2}
\end{center}
\end{figure}

Consider the set $T$ given by

\begin{equation}
  \label{eq:T}
  T = \{ \xvec(r,\yvec^{a_*}(\rho,\zvec^{a_*}))|\, r\ge \half, \rho \le \half\},
\end{equation}
where $\yvec^{a_*}(\rho,\zvec^{a_*})$ denotes the polygonal-polar coordinates
(\ref{eq:param of Chata}) on $\Chat^{a_*}$.  $T$ represents a channel in the outer
shell $\Phat(\half,1)$ through the cleaved face $\Chat^{a_*}$.  The
central axis of $T$, where $\rho = 0$, is given by $(1-r)\cvec^{a_*} + r
\pvec$, $r \ge \half$.  From (\ref{eq:marked point}) and
(\ref{eq:nhat_1}), it follows that $\nhat_1 = \svec$ along this axis.
We show that $\nhat_1$ is homotopic to a map $\nhat_2$ which is equal to $\svec$
throughout $T$, and which coincides with $\nhat_1$
for $r < \half$ and for $\yvec\notin \Chat^{a_*}$. A homotopy
$\Hhat_t(r,\yvec)$ is given by 
$\nhat_1(r,\yvec)$ for  
$r < \half$ or $\yvec\notin\Chat^{a_*}$, and for $r \ge \half$ and
$\yvec \in \Chat^{a_*}$ by
\begin{equation}
  \label{eq:n1-<n2 homotopy}
  \Hhat_t(r,\yvec^{a_*}(\rho,\zvec^{a_*})) = 
 \begin{cases}
       \nhat_1(r,\yvec^{a_*}((2\rho-t)/(2-t),\zvec^{a_*})),&t/2 < \rho \le 1,\\
    \svec,& 0\le \rho \le  t/2,
  \end{cases}
\end{equation}
where $\zvec^{a_*} \in \partial \Chat^{a_*}$.
Let $\nhat_2 = \Hhat_1$.  Then, for $r\ge \half$,
 \begin{equation}
   \label{eq:nhat_2}  
   \nhat_2(r,\yvec^{a_*}(\rho,\zvec^{a_*}))=
   \begin{cases}
     \nhat_1(r,\yvec^{a_*}(2\rho-1,\zvec^{a_*})),&\half < \rho \le 1,\\
     \svec,& 0\le \rho \le \half.
   \end{cases}
 \end{equation}
$\nhat_2$ is constant, equal to $\svec$, in the inner shell
$\Phat(\epsilon,\half)$ as well as in $T$.  See Fig.~2(c).
 
 
Next we deform $\nhat_2$ so that it is constant, equal to $\svec$,
throughout the whole inner polyhedron $\Phat(0,\half)$. This is
accomplished by displacing the polyhedral bubble in which $\nhat_1$
is varying from $\Phat(0,\epsilon)$
through the shell $\Phat(\epsilon,\half)$ and then through the channel
$T$. 
Let $\uvec$ be
parallel to the axis of $T$, ie proportional to $\cvec^{a_*}-\pvec$,
with $|\uvec|$ sufficiently large so that
\begin{equation}
  \label{eq:u condition}
  \left\{\Phat(0,\epsilon) + \uvec \right\} \cap \Phat =
\emptyset.
\end{equation}  
Choose $\epsilon$ sufficiently small so that
\begin{equation}  
  \label{eq:epsilon condition}
  \left\{\Phat(0,\epsilon)+ t \uvec \right\} \cap \Phat \subset \Phat(0,\half)\cup T, \quad 0
  \le t \le 1.  
\end{equation}  
Let
\begin{equation}
  \label{eq:n2->n3 homotopy}
  \Hhat_t(\xvec) = 
  \begin{cases}  
    \nhat_2(\xvec - t\uvec), &\xvec \in \left\{\Phat(0,\epsilon) +
    t\uvec\right\}\cap \Phat,\\
    \svec,& \xvec \in \Phat(0,\epsilon)\ \text{and}\ \xvec \notin
    \left\{\Phat(0,\epsilon) + t \uvec\right\},\\
    \nhat_2(\xvec),& \text{otherwise}.
  \end{cases}
\end{equation}  
 See Fig.~2(d).  (\ref{eq:epsilon condition}) guarantees that $\Hhat_t(\xvec)$ is
continuous, as $\nhat_2$ is continuous and is constant, equal to $\svec$, throughout
$P(\epsilon,\half)\cup T$.  Let $\nhat_3 = \Hhat_1$.  From (\ref{eq:u condition}) and
(\ref{eq:nhat_2}), it follows that $\nhat_3$ is constant, equal to
$\svec$, on $\Phat(0,\epsilon)$ and that it coincides with $\nhat_2$ in
$\Phat(\half,1)$.  See Fig.~2(e).  By applying the inverse of the  homotopy
(\ref{eq:n1-<n2 homotopy}), with $\nhat_1$ replaced by $\nhat_3$, we
can collapse the channel $T$ to  obtain a map
$\nhat_4$ (see Fig.~2(f)) given by
\begin{equation}
  \label{eq:n4}
  \nhat_4(r,\yvec) = 
  \begin{cases}
    \nihat(2r-1,\yvec),& \half \le r \le 1,\\
    \svec,& r < \half.
  \end{cases}
\end{equation}
Then
\begin{equation}
  \label{eq:n4-nihat}
  \Hhat_t(r,\yvec) = 
  \begin{cases}
    \nihat((2r -
    (1-t))/(1+t),\yvec),&\half (1-t)\le r\le 1,\\
\svec,& \rho < \half (1-t)
  \end{cases}
\end{equation}
describes a homotopy of $\nhat_4$ to $\nihat$. 
\end{proof}

\section{Surface homotopies}\label{sec: surface}

An intermediate step in the
proof of Theorem~\ref{thm: classify} is the fact that maps in
$\czero(\Phat)$ can be deformed to coincide with their
associated representative maps on $\partial \Phat$.  This is
summarised by the following:
 \begin{prop}
    \label{prop: surface homotopy}
Let $\nhat\in \czero(\Phat)$, with $\Ical = \inv(\nhat)$.
Then $\nhat$ is homotopic to a map $\nphat$ for which $\nphat =
  \nihat$ on $\partial \Phat$.
  \end{prop}
  
  To prove Proposition \ref{prop: surface homotopy}, we make use of the fact
  that deformations of $\nhat$ on the edges of $\Phat$ can be extended
  to deformations of $\nhat$ on the faces, and, similarly,
  deformations of $\nhat$ on the faces of $\Phat$ can be extended to
  deformations of $\nhat$ on $\Phat$ itself.  For completeness, we
  give an argument below which covers both cases (of course, a
  similar result holds generally on manifolds with boundary).
\begin{lem}
  \label{lem: extending boundary homotopy}
  Let $Q \subset \Rr^k$ be compact and convex with boundary $\partial
  Q$, and let $S$ be a topological space with subspace $T$.  Let
  $\czero(Q)$ denote the space of continuous maps from $Q$ to $S$ which map
  $\partial Q$ to $T$, and let $\czero(\partial Q)$ denote
  the space of continuous maps of $\partial Q$ to $T$.  Given $n \in \czero(Q)$, let
  $\partial n \in \czero(\partial Q)$ denote its restriction to
  $\partial Q$.  Suppose that $\partial n$ is homotopic to $\nu' \in
  \czero(\partial Q)$.  Then $n$ is homotopic to some $n' \in
  \czero(Q)$
  with $\partial n' = \nu'$.
\end{lem}
\begin{proof}
  Introduce polygonal-polar coordinates on $Q$.  Ie, let $\qvec$ be a
  point in the interior of $Q$, and let $\uvec(\lambda, \vvec) =
  \lambda \vvec + (1 - \lambda)\qvec$, where $0\le\lambda\le 1$ and
  $\vvec \in \partial Q$.  Given $n \in \czero(Q)$, we write, by an
  abuse of notation but for the sake of brevity, $n(\lambda,\vvec)$
  rather than $n(\uvec(\lambda,\vvec))$, and similarly for other maps
  in $\czero(Q)$.  Let $h_t$ be a homotopy from $\partial n$ to
  $\nu'$.  Let $H_t$ be given by
   \begin{equation}
     \label{eq:from boundary to interior}
     H_t(\rho,\vvec) = 
     \begin{cases}
       h_{2\rho+t-2}(\vvec),&1 - \half t < \rho \le 1,\\
       n(\rho/(1-\half t),\vvec),&\rho \le 1 - \half t.
     \end{cases}
   \end{equation}
Let $n' = H_1$.  Then $n$ is homotopic to $n'$, and $\partial n' =
\nu'$.
\end{proof}

\begin{proof}[Proof of Proposition~\ref{prop: surface homotopy}]
  
  Let $\czero(\partial \Phat)$ denote the space of continuous tangent
  unit-vector fields on the boundary of $\Phat$ (so that
  $\nhat(\yvec)$ is tangent to $\partial \Phat$ at $\yvec$).  
  Given $\nhat\in \czero(\Phat)$, let $\partial \nhat \in
  \czero(\partial\Phat)$ denote its restriction to $\partial \Phat$.

  From Lemma \ref{lem:
    extending boundary homotopy}, it suffices to show that
\begin{equation}
  \label{eq:boundary homotopy}
  \partial \nhat \sim \partial \nihat,
\end{equation}
where we have the usual notion of homotopic equivalence in
$\czero(\partial \Chat)$. 
We establish (\ref{eq:boundary homotopy}) in two steps, first deforming $\partial \nhat$ to
coincide with $\partial \nihat$ on the edges
of $\partial\Phat$, and then deforming it
further to coincide with $\partial \nihat$ on the faces of $\partial \Phat$.

Since $\nhat$ and $\nihat$ have the same edge orientations
(ie, $\evec^b(\nhat) = \epsilon^b$), they coincide on truncated edges, and
therefore coincide on the endpoints of the cleaved edges $\Bhat^{ac}$.
Since $\nhat$ and $\nihat$ have the same kink numbers, there is a
homotopy between the restrictions of $\nhat$ and $\nihat$ 
to the cleaved edges. (Explicitly, if, on $\Bhat^{ac}$,
$\nhat$ is represented by an angle $\theta^{ac}(s)$ in the plane
tangent to $\Fhat^c$, with $0\le s \le 1$, and $\nihat$ is similarly
represented by $\theta'^{ac}(s)$ with $\theta'^{ac}(0) =
\theta^{ac}(0)$, then $k^{ac} = \kappa^{ac}$ implies that
$\theta'^{ac}(1) = \theta^{ac}(1)$, and a homotopy is given by
$(1-t)\theta^{ac}(s) + t\theta'^{ac}(s)$).
By Lemma~\ref{lem: extending boundary homotopy}, these homotopies on
$\Bhat^{ac}$ can be extended to homotopiwhich will be convenient in what follows.

es on the faces of $\Phat$,
and therefore to a homotopy $\hhat_t$ on $\partial \Phat$.  Let
$\nuphat = \hhat_1$.  By construction, $\nuphat$ coincides with
$\partial \nihat$ on the edges of $\partial\Phat$.

Next, we construct homotopies from $\nuphat$ to $\partial \nihat$ on
the faces of $\Phat$.  On the truncated face $\Fhat^c$, $\nuphat$ may
be represented by an angle $\theta'^c(\yvec^c)$ in the plane tangent to
$\Fhat^c$.  The sum rule (\ref{eq:kink sum rule}) ensures that
$\theta'^c(\yvec^c)$ is continuous.  $\partial \nihat$ may be similarly
represented by $\theta^c(\yvec^c)$.  By construction,
$\theta'^c(\yvec^c)$ and $\theta^c(\yvec^c)$ agree on $\partial
\Fhat^c$ up to addition of a multiple of $2\pi$, which we can take to
vanish.  A homotopy between them on $\Fhat^c$ is given by $(1-t)\theta'^c(\yvec^c) +
t\theta^c(\yvec^c)$.

Homotopies from $\nuphat$ to $\partial \nihat$ on the cleaved faces
may be constructed as follows.  Let $\yvec^a(\rho,\zvec^a)$ be the
polygonal-polar coordinates on $\Chat^a$ given by (\ref{eq:param of
  Chata}), with $0\le \rho \le 1$ and $\zvec^a \in \partial
\Chat^a$.  We first deform $\nuphat$ so that it agrees with $\partial \nihat$
for $\rho \ge \half$.  A homotopy is given by 
\begin{equation}
  \label{cleave homotopy}
  \hhat^a_t(\rho,\zvec^a) = 
  \begin{cases}
     \partial \nihat(2\rho - 1,\zvec^a),& 1-\half t< \rho \le 1,\\
     \partial \nihat(5-4\rho - 3t,\zvec^a),& 1-\thfourth t < \rho \le
     1-\half t,\\
 \nuphat(\rho/(1-\thfourth t),\zvec^a),& 0\le \rho \le 1-\thfourth t.
  \end{cases}
\end{equation}
Let $\nupphat = \hhat^a_1$.  Then $\nupphat$ coincides with $\nihat$ for
$\rho \ge \half$.  

The region $\rho\le \half$ on $\Chat^a$ is a topological two-disk.  On
the boundary, where $\rho = \half$, $\nupphat$ and $\partial\nihat$ are
both constant, equal to $-\svec$ (cf (\ref{eq:nihat, rho > half}) and
(\ref{eq:nhat^a_0}).)
By identifying points on
the boundary, we may regard $\nupphat$ and $\nihat$ as
maps on $S^2$ which preserve a marked point $-\svec$.
The fact that $w^a(\nihat) = \omega^a$ implies that these maps have the same
degree, and therefore are homotopic.  Thus there exists a homotopy on
$\rho \le \half$ which takes $\nupphat$ to $\partial \nihat$ and which is equal
to $-\svec$ for $\rho = \half$.  This establishes a homotopy between
$\nupphat$ and $\partial \nihat$ on $\Chat^{a}$.

Together, the homotopies on truncated faces and cleaved faces give a  homotopy
from $\nupphat$ to $\partial \nihat$.  The chain of equivalences $\partial
\nhat \sim \nuphat \sim \nupphat \sim \partial \nihat$ in
$\czero(\partial \Phat)$ gives the required result.

\end{proof}

\section{Concluding remarks}
The problem considered here may be generalised to $n > 3$ dimensions.
Generalisations suggested by liquid crystal applications include
normal boundary conditions (ie, on the faces of $P$, $\nvec$ is
required to be orthogonal to the faces), and periodic boundary
conditions on a cubic domain from which a polyhedral domain has been
excised (this corresponds to an array of liquid crystal cells with
polyhedral geometries).  It may be our results apply to nonconvex
polyhedra as well.

We thank Adrian Geisow and Chris Newton for stimulating our interest
in this area, and Apala Majumdar for helpful comments.

\bibliography{classnotes13}

\end{document}